\documentstyle[12pt,aaspp4]{article}

\begin{document}

\vspace*{-0.5in}
\hfill{Accepted for publication in the Astrophysical Journal Letters}

\title {Ultraviolet Galaxy Counts from STIS Observations of the
Hubble Deep Fields\footnote[1]{Based on observations made with the
NASA/ESA {\em Hubble Space Telescope}, obtained from the Space
Telescope Science Institute, which is operated by the Association
of Universities for Research in Astronomy, Inc., under NASA contract
NAS 5-26555.}.}

\author {Jonathan P. Gardner\altaffilmark{2}, 
Thomas M. Brown\altaffilmark{2,4},
Henry C. Ferguson\altaffilmark{3},
\\
\altaffiltext{2}{Laboratory for Astronomy and Solar Physics,
Code 681, Goddard Space Flight Center, Greenbelt MD 20771}
\altaffiltext{3}{Space Telescope Science Institute,
3700 San Martin Drive, Baltimore MD 21218}
\altaffiltext{4}{NOAO Research Associate}
}

\begin{abstract}

We present galaxy counts in the near-ultraviolet (NUV; 2365\AA)
and far-ultraviolet (FUV; 1595\AA) obtained from Space Telescope
Imaging Spectrograph (STIS) observations of portions of the Hubble
Deep Field North (HDFN), the Hubble Deep Field South (HDFS), and
a parallel field near the HDFN. All three fields have deep (AB$>$29)
optical imaging, and we determine magnitudes by taking the ultraviolet
flux detected within the limiting optical isophote. An analysis of
the UV-optical colors of detected objects, combined with a visual
inspection of the UV images, indicates that there are no detectable
objects in the UV images that are not also detected in the optical.
We determine our detection area and completeness as a function of
magnitude with a simulated distribution of galaxies based on the
HDFN Wide Field Planetary Camera 2 (WFPC2) V+I image. The galaxy
counts reach to $NUV_{AB}=29$ and $FUV_{AB}=30$, which is 1 magnitude
fainter than the HDFN WFPC2/F300W counts, and 7 magnitudes fainter
than balloon-based counts. We integrate our measurements to present
the extragalactic background radiation coming from resolved objects.
The NUV galaxy counts show a flattening or turnover beginning at
about $NUV_{AB}=26.0$, which is not predicted either by no-evolution
models based upon a local luminosity function with a steep faint
end slope, nor by a semi-analytic model in which starbursts are
caused by major mergers. The FUV counts also show a flat slope. We
argue that the flat slopes could be caused by a short duty cycle
for star formation, additional starbursts triggered by minor mergers,
and an extended quiescent phase between starburst episodes.

\end{abstract}


\section{Introduction}

Imaging of random fields at high galactic latitude is a powerful
tool for investigating galaxy formation and evolution. Number counts
of galaxies, as a function of filter bandpass, have provided the
first solid evidence for the evolution of the normal galaxy population
(e.g., Tyson \markcite{tyson88}1988; Lilly, Cowie \& Gardner \markcite{lilly91}1991). Much work has been done in
the optical and near-infrared, and the number counts of galaxies
are now well established observationally at bright magnitudes
through wide-area surveys done both with electronic detectors
(Gardner et al.\ \markcite{gardner96}1996) and photographic plates (Maddox et al.\ \markcite{maddox90}1990). At
faint magnitudes, the Hubble Deep Field projects have established
the galaxy counts in the optical and near-infrared to AB=30 magnitude
(Williams et al.\ \markcite{williams96}1996; Thompson et al.\ \markcite{thompson99}1999; Gardner et al.\ \markcite{gardner00}2000;
Ferguson, Dickinson \& Williams \markcite{ferguson00}2000). The AB magnitude system is defined by $AB =
-2.5log(F_{\nu}) - 48.6$ where $F_{\nu}$ is given in
erg~cm$^{-2}$~s$^{-1}$~Hz$^{-1}$ (Oke \protect\markcite{oke71}1971).

Bright galaxy counts in the ultraviolet have been measured from
balloon-borne imaging in the 2000{\AA} window by the FOCA experiment
(Milliard et al.\ \markcite{milliard92}1992), with follow-up ground-based optical spectroscopy
by Treyer et al.\ \markcite{treyer98}(1998) and Sullivan et al.\ \markcite{sullivan00}(2000). With a 40 cm telescope,
FOCA surveyed a total area of about 6 square degrees near the North
Galactic Pole over the magnitude range $17.75<F(2000)_{AB}<20.75$,
measuring a nearly Euclidean slope for the counts. The colors and
spectra of their detected galaxies are consistent with an actively
star-forming population dominated by Sdm or later types in the
redshift range $0<z<0.5$. The data show a larger population of
UV-emitting galaxies than had previously been assumed (Armand \& Milliard \markcite{armand94}1994),
with a steep faint-end slope for the UV luminosity function.

In this Letter, we present faint galaxy counts in the near- and
far-ultraviolet (NUV \& FUV) obtained by imaging several fields
with the Space Telescope Imaging Spectrograph (STIS; Kimble et al.\ \markcite{kimble98}1998;
Woodgate et al.\ \markcite{woodgate98}1998).

\section{Description of the data}

We have examined three datasets. The largest field consists of
follow-up STIS imaging of part of the Hubble Deep Field North
(HDFN-Fol; Williams et al.\ \markcite{williams96}1996), and contains 6 fields of view with
the FUV and NUV detectors. The second field consists of the primary
STIS observations of the Hubble Deep Field South (HDFS-Pri;
Gardner et al.\ \markcite{gardner00}2000). The third field was a parallel field imaged with
STIS in the optical and ultraviolet during a deep imaging campaign
of the HDFN (hereafter HDFN-Par) done with NICMOS (Thompson et al.\ \markcite{thompson99}1999).
It is located $\sim8.5\arcmin$ East of the deep field. All observations
used the F25QTZ filter, and are dark noise limited in both the FUV
and NUV. These data have been used for a measurement of the diffuse
FUV background emission (Brown et al.\ \markcite{brown00}2000). Details of the field and
the FUV exposure times are given in that paper. The NUV exposure
times range from 16,865s to 22,616s per field of view.

We used the public reduction of the HDFS-Pri field (Gardner et al.\ \markcite{gardner00}2000),
and reduced the data on the HDFN-Fol and HDFN-Par fields in the
same way. All of the exposures are dithered, which results in a
variable exposure time across the final images. In addition, the
presence of the FUV dark glow (described in detail by Gardner et al.\ \markcite{gardner00}2000
and Brown et al.\ \markcite{brown00}2000) results in noise that varies by a factor of 2
across the field.

All three fields have deep optical imaging, reaching a 3$\sigma$
limiting AB magnitude of $\gtrsim$29. The HDFN-Fol and HDFS-Pri
fields are contained within the deep optical fields. The HDFN-Par
field was observed with STIS in the 50CCD mode for 34,076 s. The
HDFN-Par observations were made during the first NICMOS camera 3
campaign when the focus of the telescope was changed, and are
therefore slightly out of focus, and somewhat less sensitive to
point sources than they would otherwise be. However, we do not make
a distinction between point sources and extended sources in our
analysis, and we take source size into account in our determination
of the galaxy counts, so this will not affect our results.

\subsection{Object detection and photometry}

We used the publicly released {\sc SExtractor} (Bertin \& Arnouts \markcite{bertin96}1996)
version 2 catalog for the HDFS-Pri field, and followed a similar
procedure for object detection and photometry on the other two
fields. As {\sc SExtractor} has problems handling quantized low-signal
data, we computed the flux in the UV images within the pixels
assigned to each object in the optical, with local background
subtraction. See Gardner et al.\ \markcite{gardner00}(2000) for a more detailed discussion
of this procedure. We have made the assumption that there are no
objects detectable on the UV images that are undetected on the
optical images. Visual inspection of the images confirms this
assumption. In addition, the photometry are isophotal magnitudes,
where the isophote has been determined in the optical.

\begin{figure}

\plotone{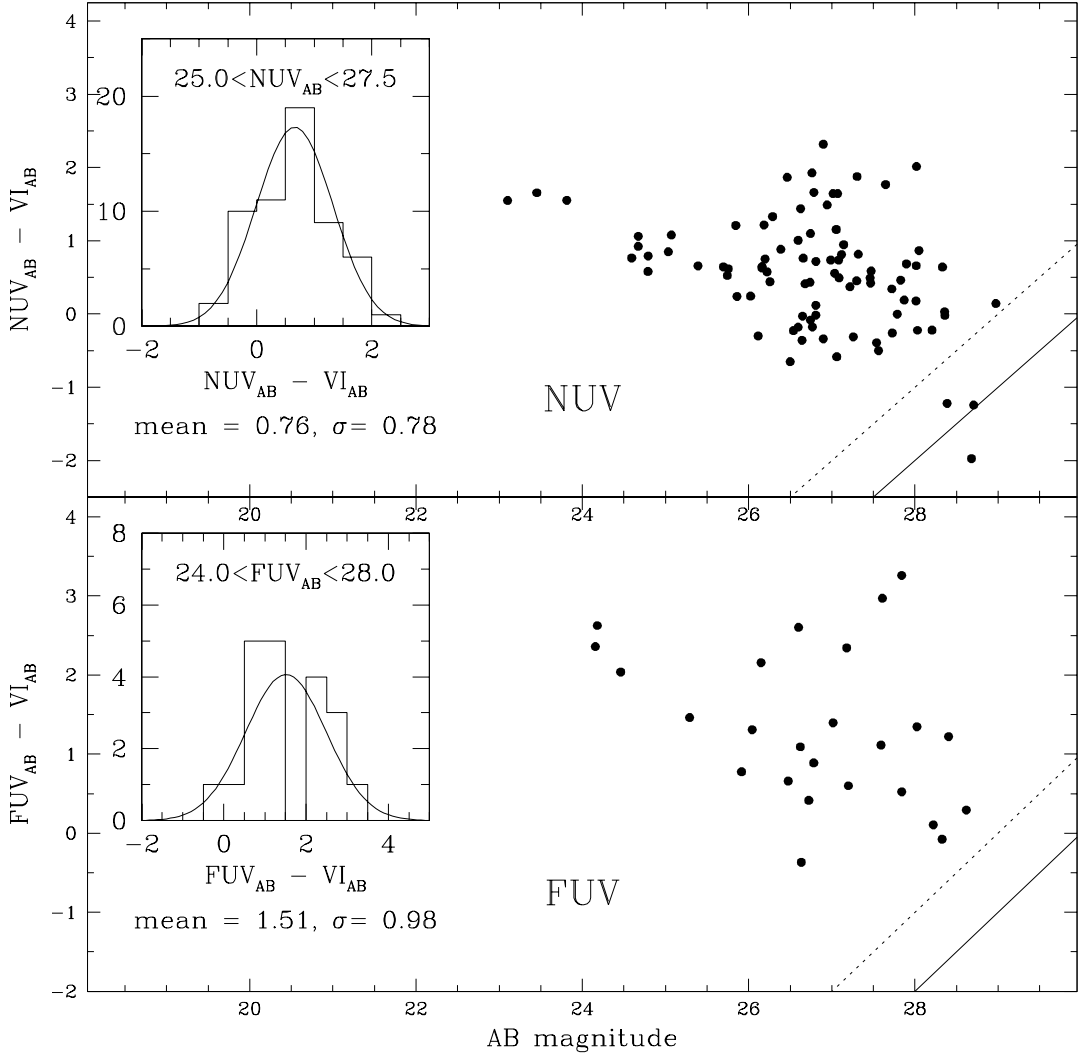}

\caption{Ultraviolet to optical colors of detected galaxies. In
the top panel we plot the $NUV_{AB}-VI_{AB}$ colors of galaxies from
the HDFN-Fol field as a function of $NUV_{AB}$. The dotted line
represents $VI_{AB} = 29$, the nominal HDFN completeness limit,
while the solid line represents $VI_{AB}=30$. While there are three
galaxies fainter than the HDFN optical completeness limit, they
are in the faintest NUV magnitude bin. Our procedure of measuring
UV fluxes only within isophotes detected in the optical is unlikely
to be incomplete at $NUV_{AB}<28$. Visual inspection of the UV
images does not reveal any objects detected in the UV that are not
also detected in the optical. As an inset to the top panel, we plot
a histogram of the colors in the range $25<NUV_{AB}<27.5$, which
we used for our completeness determination. The bottom panel contains
the same information for the FUV. There are no galaxies detected
in the FUV beyond the HDFN optical completeness limit.}

\label{colmag}

\end{figure}

In Figure~\ref{colmag} we plot the UV-optical colors of the detected
galaxies in the HDFN-Fol field. The optical flux is determined from
the {\sc SExtractor} version 2 catalog of the sum of the HDFN F814W
and F606W images (hereafter VI). We plot as a solid line the location
on this figure where objects with VI$_{AB}$=30 would be, and as a
dotted line where objects with VI$_{AB}$=29 would be. If we take
$VI_{AB}$=29 as the completeness limit for the HDFN catalog, then
there are only three galaxies detected at the $3\sigma$ limit in
our NUV catalog and no galaxies in our FUV catalog that are fainter
than this in the optical. The three galaxies are fainter than
$NUV_{AB}$=28 in the UV, and the UV-optical color distribution does
not show a sharp cut-off at the optical detection limit.

\subsection{Measurement of the number counts}

There has been considerable debate about incompleteness in measurements
of galaxy counts due to the distribution of central surface brightness
in galaxies (McGaugh \markcite{mcgaugh94}1994; Ferguson \& McGaugh \markcite{ferguson95}1995). For every
photometric survey the apparent magnitude detection limit is a
function of central surface brightness, or galaxy size
(Petrosian \markcite{petrosian98}1998). While this criticism has been directed
at surveys done with photographic plates and bright isophotal
detection limits, the high resolution of HST data makes them less
sensitive to extended objects. We determine the area and the
completeness for each apparent magnitude galaxy bin simultaneously,
using simulations which sample the optical HDFN catalog.

\begin{figure}

\plotone{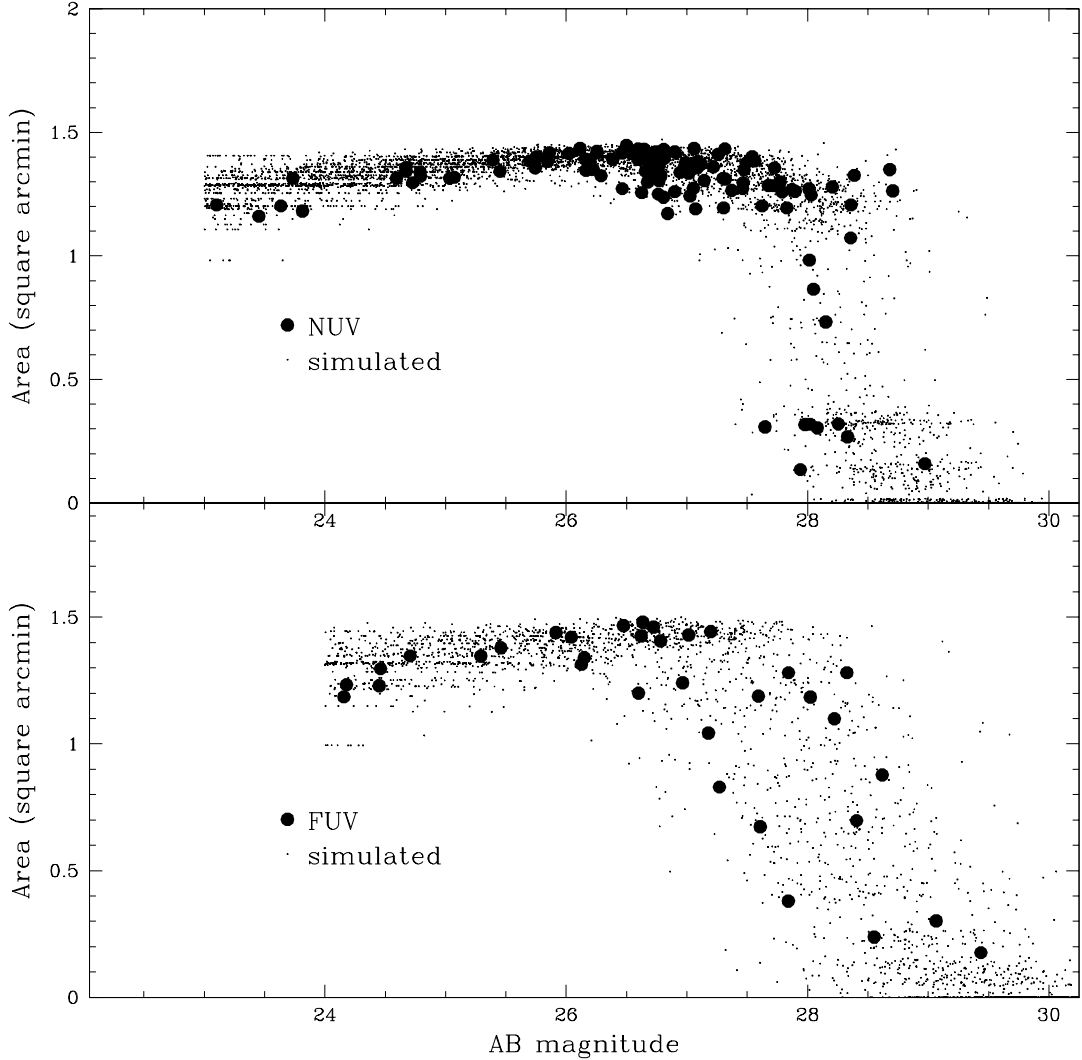}

\caption{Detection areas for observed and simulated galaxies. For
each detected galaxy and for those in our simulations, we calculated
the total area over which we would have detected the galaxy at the
$3\sigma$ level. This area for each galaxy depends both on its
magnitude and on its size. For each galaxy, we excluded an area
around the perimeter of each field, and the slope at bright magnitudes
is due to these edge effects. The simulated galaxies are drawn from
the HDFN V+I catalog, and the horizontal lines at bright magnitudes
are due to repeats. The horizontal layers at faint magnitudes in
the NUV are due to the fact that our three fields have different
detection limits. Because of the presence of the dark current glow,
the depth of the FUV images is smoothly variable over the detector, and
the horizontal lines are less visible in that section of the Figure.}

\label{areafig}

\end{figure}

We use the error maps to determine the area over which each galaxy
would have been detected. We smooth them with a median filter of
$0.4\arcsec \times 0.4\arcsec$, in order to remove small-scale
variations introduced during hot pixel removal or the drizzling
process. The number of pixels where a galaxy of a given size and
magnitude would be detected at 3 sigma gives the area. To eliminate
edge effects, we remove a region around the perimeter of each field
equal to the radius of a circle with the same area as the galaxy.
We plot these areas as a function of magnitude in Figure~\ref{areafig}.
At bright magnitudes, the galaxies would be detected anywhere in
our survey area, except the edges of the images. At fainter levels,
the detection area depends on galaxy size, and drops to zero where
galaxies are too faint to be detected at all.

Next we construct a catalog of simulated galaxies to correct for
incompleteness in our measured counts, using the isophotal sizes
and optical magnitudes from the entire HDFN VI catalog. We simulate
400 galaxies in each UV magnitude bin. We convert these to simulated
VI magnitudes with random UV-optical colors selected from a Gaussian
distribution plotted as an inset to Figure~\ref{colmag}. We then
take from the HDFN catalog the galaxy with the closest-matching
optical magnitude. This gives us a size for the simulated galaxy,
effectively bootstrap sampling the size distribution of the HDFN
VI catalog, but using a UV-selected galaxy sample. For each simulated
galaxy, we determine the area of detection, and plot this in
Figure~\ref{areafig}. In each of our apparent magnitude bins, the
combined area and completeness is the average of all of the detection
areas for the simulated galaxies, including those simulated galaxies
for which the detection area is zero.

We make no attempt at star-galaxy separation, other than to remove
the bright point source in the middle of the HDFN-Fol field. The
region around and including the target quasar has also been removed
from the HDFS-Pri field.

\subsection{Isophotal to total magnitude correction}

We plot the galaxy counts in Figure~\ref{uvnc}, and list them in
Table~\ref{counts}. The data plotted in Figures~\ref{colmag}
and~\ref{areafig} refer to the isophotal magnitudes. We have also
corrected the isophotal magnitudes to total magnitudes using
simulations of the object detection procedure in the HDFN VI image,
as described by Ferguson \markcite{ferguson98}(1998). We fit a 5th order polynomial
to the median difference of the isophotal and total magnitudes in
the simulations in half-magnitude bins, as a function of isophotal
magnitude. The corrections are $\leq 0.20$ mag for the galaxies at
$VI_{AB}<28$, the majority of our sample. This procedure operates
on the assumption that there are no UV-to-optical color gradients
in the galaxies, a reasonable approximation for spiral galaxies.
Elliptical galaxies tend to have larger isophotal-to-total corrections
in the optical than spiral galaxies, but are more centrally
concentrated in the UV than in the optical. These corrections tend
to steepen the counts slightly (every galaxy is made brighter),
but the changes are mostly within the Poissonian errors on the
bins.

\subsection{Extragalactic background radiation from resolved objects}

Brown et al.\ \markcite{brown00}(2000) measured the diffuse FUV background radiation
after removing the resolved objects. The contribution of
resolved galaxies is the integral of the flux from the
galaxy counts; this flux is dominated by galaxies near the break where
K-corrections become significant. Since we do not sample this break,
we construct a model of the counts extending to bright levels. We
convert the FOCA counts to the FUV using the model of Gardner \markcite{gardner98}(1998),
which gives $F(2000)_{AB} - FUV_{AB} = 0.05$ for star-forming
galaxies at low redshift. We fit a power law with a Euclidean slope
of 0.6 to these data (intercept $-9.9\pm0.15$), and fit a power
law to our data (slope $0.14\pm0.06$, intercept $0.7\pm1.6$).  The
integrated background from this model is $\nu I_{\nu} =
3.9^{+1.1}_{-0.8}~nW~m^{-2}~sr^{-2}$, or $I_{\lambda} =
195^{+59}_{-39}~ph~s^{-1}~cm^{-2}~sr^{-1}~{\AA}^{-1}$ at 1595{\AA},
where the statistical errors come from the errors in the
fitted slopes. This is an upper limit, and as an alternative,
we fit a third power laws connecting the faintest FOCA point with
our brightest point (slope $0.45\pm0.06$, intercept $-7.0\pm1.3$).
This gives a lower limit of $2.9^{+0.6}_{-0.4}$ in sterance, or
$144^{+28}_{-19}$ in photon units. Our range of 144 to 195 photon
units is mildly inconsistent with the prediction by Armand, Milliard, \& Deharveng \markcite{armand94.2}(1994)
of 40 to 130 photon units. Of the $\sim$300 photon units measured
for the total background (Bowyer \markcite{bowyer91}1991), our measurements reduce
the contribution required from truly diffuse sources (e.g.,
dust-scattered Galactic starlight, the inter-galactic medium, or
\ion{H}{2} two-photon emission).

We fit our NUV data with two power laws, breaking at
$NUV_{AB}=26.75$, and we convert the FOCA counts to the NUV using
$F(2000)_{AB}-NUV_{AB}=-0.02$. The three power laws are (slope,
intercept) = ($0.6, -10.0\pm0.15$), ($0.33\pm0.07$,$-3.7\pm1.7$),
and ($-0.06\pm0.09$, $6.7\pm2.5$). The integrated background is
$3.6^{+0.7}_{-0.5}$ in sterance, or $179^{+35}_{-25}$ in photon
units.

\section{Discussion}

The most striking feature of the faint counts is the very flat
slope of the NUV counts at the faintest magnitude bins. In the
range $26.5<NUV_{AB}<29.0$, the slope is $-0.06\pm0.09$, formally
a turnover. Changes in the slope of the galaxy counts can
contain information about the cosmological parameters (Yoshii \& Peterson \markcite{yoshii91}1991).
The intrinsic and intergalactic absorption at wavelengths shorter
than the Lyman limit, which removes nearly all flux, provides a
redshift cut-off for galaxies in the observed UV, and counts fainter
than $M_*$ at this redshift are effectively volume-limited, sampling
the faint end of the luminosity function. The flat slope we see is
inconsistent with a no-evolution model extrapolation of the luminosity
function measured locally by Sullivan et al.\ \markcite{sullivan00}(2000), indicative of
evolution in the star formation properties of galaxies (see also
Steidel et al.\ \markcite{steidel99}1999).

\begin{figure}

\plotone{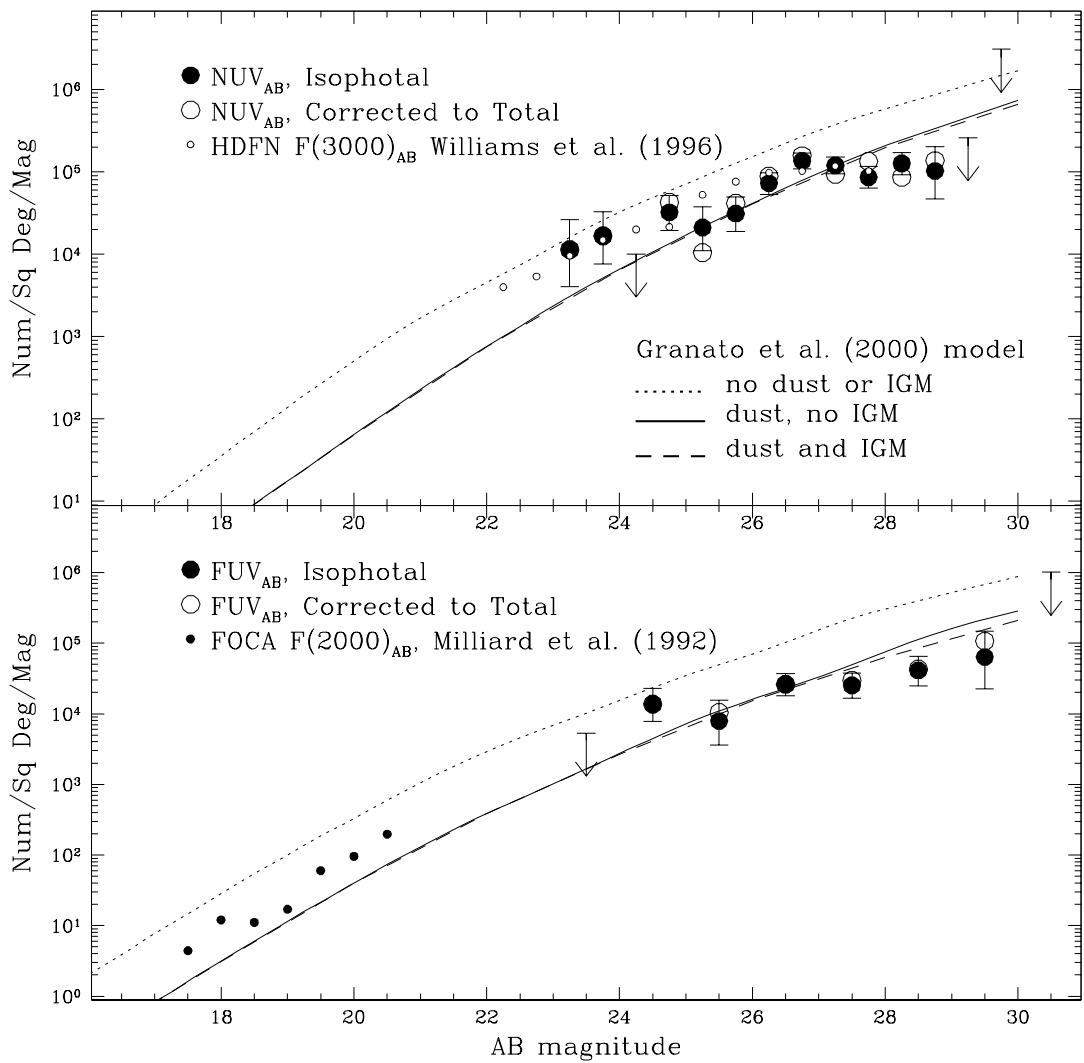}

\caption{Galaxy counts. The large solid symbols are the isophotal
counts, corrected for completeness as described in the text. The large
open symbols have the magnitudes corrected to total. In the upper
section of the Figure, we plot the WFPC2/F300W counts from
Williams et al.\ \protect\markcite{williams96}(1996) and in the lower section we plot the
balloon-based FOCA/2000{\AA} counts of Milliard et al.\ \protect\markcite{milliard92}(1992) for
comparison. All counts are plotted in AB magnitudes, but we have
not attempted to correct the comparison counts for the different
bandpasses.}

\label{uvnc}

\end{figure}

In Figure~\ref{uvnc} we plot a prediction of the UV galaxy
counts from the model of Granato et al.\ \markcite{granato00}(2000). The prediction is much
steeper than the observed counts, underpredicting them at bright
magnitudes and overpredicting the number of faint galaxies. This
model assumes that the rate of star formation is proportional to
the rate of infalling cold gas, unless a major merger takes place.
In a merger between galaxies with a mass ratio of 0.3 or greater,
all of the available gas is converted into stars. A possible
explanation for the flat slope of the counts is given by
Somerville, Primack \& Faber \markcite{somerville00}(2000), based on a suggestion by White \& Frenk \markcite{white91}(1991). They
investigate a scenario in which all mergers, independent of mass
ratio, trigger starbursts. Additional starbursting galaxies (within
the context of models in which the total starformation rate is
fixed), have the effect of increasing the number of bright galaxies
and decreasing the number of faint galaxies in the luminosity
function. Additional flattening of the counts will be seen if the
duty cycle for starbursts is short, and is followed by an extended
quiescent phase between the starburst episodes.

\acknowledgments

We would like to acknowledge the contributions of our co-investigators
on the HDFN-Fol project: Mark Dickinson, Andy Fruchter, Mauro
Giavalisco, Sally Heap, Ray Lucas, Piero Madau, Eliot Malumuth,
Lucia Pozzetti, Bruce Woodgate and Gerry Williger. We would like
to thank Carlos Frenk, Cedric Lacy, Simon White and Rachel Somerville
for useful discussions and model predictions. We would like to
thank Rodger Thompson, Nick Collins and Robert S. Hill for assisting
with the HDFN-Par field. We would like to thank Bob Williams and
the HDFN and HDFS teams. Support for this work was provided by NASA
through grants numbered GO-07410.03-96A and AR-08380.01-97A from
the Space Telescope Science Institute which is operated by the
Association of Universities for Research in Astronomy, Inc., under
NASA contract NAS5-26555. TMB acknowledges support from the STIS
Investigation Definition Team and the Goddard Space Flight Center.

\clearpage

{\scriptsize
\begin{deluxetable}{lrrrrrrrr}
\tablecaption{Galaxy Counts}

\tablehead{
\colhead{ABmag} &
\multicolumn{5}{c}{Isophotal Magnitude} &
\multicolumn{3}{c}{Total Magnitude} \\
\colhead{} &
\colhead{Raw N} &
\colhead{$log(N)$} &
\colhead{$\sigma_{low}$} &
\colhead{$\sigma_{high}$} &
\colhead{area} &
\colhead{Raw N} &
\colhead{$log(N)$} &
\colhead{area}
}

\startdata
FUV&&&&&&&&\nl
 23.50&  0&  \nodata &  \nodata &  3.73&  1.25&  0&  \nodata &  1.26\nl
 24.50&  5&  4.13&  0.25&  0.22&  1.32&  5&  4.14&  1.31\nl
 25.50&  3&  3.90&  0.34&  0.30&  1.37&  4&  4.02&  1.37\nl
 26.50& 10&  4.42&  0.16&  0.15&  1.38& 10&  4.42&  1.37\nl
 27.50&  8&  4.40&  0.18&  0.17&  1.14&  9&  4.47&  1.09\nl
 28.50&  6&  4.61&  0.22&  0.20&  0.53&  5&  4.63&  0.42\nl
 29.50&  2&  4.80&  0.45&  0.37&  0.11&  1&  5.03&  0.03\nl
 30.50&  0&  \nodata &  \nodata &  6.00&  0.01&  0&  \nodata &  0.00\nl
\nl
NUV&&&&&&&&\nl
 23.25&  2&  4.06&  0.45&  0.37&  1.27&  2&  4.05&  1.27\nl
 23.75&  3&  4.22&  0.34&  0.30&  1.30&  3&  4.22&  1.30\nl
 24.25&  0&  \nodata &  \nodata &  4.00&  1.32&  0&  \nodata &  1.32\nl
 24.75&  6&  4.51&  0.22&  0.20&  1.35&  8&  4.63&  1.35\nl
 25.25&  4&  4.32&  0.28&  0.25&  1.37&  2&  4.02&  1.38\nl
 25.75&  6&  4.49&  0.22&  0.20&  1.39&  8&  4.62&  1.39\nl
 26.25& 14&  4.86&  0.13&  0.13&  1.39& 17&  4.94&  1.39\nl
 26.75& 26&  5.13&  0.09&  0.09&  1.38& 30&  5.20&  1.38\nl
 27.25& 22&  5.08&  0.10&  0.10&  1.33& 17&  4.97&  1.31\nl
 27.75& 14&  4.93&  0.13&  0.13&  1.17& 20&  5.13&  1.08\nl
 28.25& 13&  5.10&  0.14&  0.13&  0.74&  5&  4.93&  0.42\nl
 28.75&  3&  5.01&  0.34&  0.30&  0.21&  1&  5.14&  0.05\nl
 29.25&  0&  \nodata &  \nodata &  5.41&  0.05&  0&  \nodata &  0.00\nl
 29.75&  0&  \nodata &  \nodata &  6.49&  0.004&  0&  \nodata &  0.00\nl

\tablecomments{Magnitudes represent the center of the bins, counts
are given in $log_{10}$(N mag$^{-1}$ deg$^{-2}$), errors are
Poisonnian and are taken from the calculations of Gehrels \protect\markcite{gehrels86}(1986),
and detection areas are given in square arcminutes.}

\label{counts}
\enddata
\end{deluxetable}
}

\clearpage

\end{document}